\documentclass[prd,twocolumn,superscriptaddress]{revtex4}

\usepackage{graphicx}
\usepackage{hyperref}

\renewcommand{\slash}{\not}

\makeatletter
\@addtoreset{equation}{section}

\makeatother

\newcommand{\beq}{\begin{equation}}
\newcommand{\eeq}{\end{equation}}

\newcommand{\di}{\displaystyle}

\newcommand{\ga}{\gamma}
\newcommand{\Ga}{\Gamma}
\newcommand{\la}{\lambda}

\newcommand{\si}{\sigma}
\newcommand{\ve}{\varepsilon}

\newcommand{\pk}{{k \cdot p}}
\newcommand{\pkprime}{{k' \cdot p}}

\newcommand{\pq}{{q\cdot p}}
\newcommand{\W}{{p'}}
\newcommand{\qW}{{q \cdot p'}}

\newcommand{\pW}{{p\cdot p'}}

\newcommand{\GeV}{\; {\mathrm{GeV}}}
\newcommand{\cm}{\; {\mathrm{cm}}}

\begin{document}

\title{
Resonance production by neutrinos: I. $J=3/2$ Resonances.
}
\author{Olga Lalakulich}
\affiliation{Institut of Physics, Dortmund University, 44221, Germany}
\email[E-mail:]{olalakul@zylon.physik.uni-dortmund.de}
\author{Emmanuel A. Paschos}
\affiliation{Institut of Physics, Dortmund University, 44221, Germany}
\affiliation{Fermi National Laboratory, Batavia, IL, 60510}
\email[E-mail:]{paschos@physik.uni-dortmund.de}
\date{\today}

\begin{abstract}
The article contains general formulas for the production of $J=3/2$ resonances by neutrinos and antineutrinos. It specializes to the $P_{33}(1232)$ resonance whose form factors are determined by theory and experiment and then are compared with experimental results at low and high energies. It is shown that the minimum in the low $Q^2$ region is a consequence of a combined effect from the vanishing of the vector form factors, the muon mass and Pauli blocking. Several improvements for the future investigations are suggested.
\end{abstract}


\maketitle

\section{Introduction}

One pion production is a process, which, along  with the quasi-elastic scattering and DIS, contribute to the total cross section of neutrino interactions with nuclei. If the pion is absorbed in the nuclear medium of the target, this process constitutes the major background to the quasi-elastic process. It is known that the one pion production proceeds through resonance production, the leading contribution coming from $\Delta^{++}$ resonance. In the coming era of long baseline neutrino experiments the cross section of this process needs to be calculated with  high accuracy. At the same time one must identify and separate the coherent component.

Theoretical and experimental study of one-pion neutrino production was performed in 70's.  A comprehensive experimental study, including the $Q^2-$de\-pen\-dence of the differential neutrino cross section, was made in two experiments at ANL and BNL. Below we represent the data from these two and other experiments and fit them with theoretical formulas.

Theoretically the model of Rein and Segal \cite{Rein:1980wg} is often used to estimate the neutrino production of the resonances. This model is based on the quark harmonic oscillator model for the form factors developed by Feynman, Kislinger and Randal \cite{Feynman:1971wr}.  Another approach is to parametrize the neutrino-nucleon-resonance vertex with phenomenological form factors. The cross section is usually expressed in terms of helicity amplitudes \cite{Zucker:1971hp,LlewellynSmith:1971zm} with detailed formulas given in Ref. \cite{Schreiner:1973mj}. After the discovery of neutrino oscillations the production of resonances by muon- amd tau-neutrinos was studied again \cite{Paschos:2000be,Paschos:2001np,Yang:1998zb,Paschos:2003qr,Alvarez-Ruso:1998hi} as a means for extracting oscillation parameters.  The calculations have been done, neglecting the muon mass, which is a valid approximation for $Q^2 >> m_\mu^2$. 

Since the time of the ANL and BNL experiments, it is known that there is a difference between the data and theoretical predictions in the region of small $Q^2$ ($Q^2<0.1\GeV^2$). Nowadays it appears that the same problem revealed itself in new experiments (K2K and MiniBooNE). In this region of $Q^2$ the results can be influenced by the muon mass, Pauli blocking and coherent pion production. In this article we take into account the nonzero muon mass and Pauli blocking and calculate the cross sections independently.  We decided to calculate the cross section, making use of the phenomenological form factors. Numerical and analytical comparisons show, that our results agree with the standard formulas. 

The formulas we present here follow closely the notation from deep inelastic scattering where the cross section is given in terms of structure fuctions ${\cal W}_i(Q^2,\nu)$, with the leptonic variables occuring in multiplicative factors. The mass of the muon occurs in the multiplicative factors and also enters indirectly when we define the accessible region of phase space. We compare our results with the production of $\Delta^{++}$ resonance, where only the amplitude with isospin $3/2$ contributes.

The plan of the paper is as follows. In section \ref{resprod} we collect general formulas for the production of resonances with $J=3/2$. These formulas, together with those in Appendix, enable the reader to write a program and produce the cross section. We give values for the coupling constants and form factors wich are frequently used. In section \ref{experresults} the production of the $P_{33}(1232)$ resonace is compared to available data, including muon mass effects. Finally, in sections \ref{specialprop} and \ref{summaryconclu} we highlight special properties and point out interesting features to be investigated in the experiments.

\section{Resonance production\label{resprod}}

In this article we discuss experiments in which the reaction 
\beq
\nu(\vec k) \, p(\vec p) \to \mu^-(\vec k')\, \Delta^{++}(\W) \to \mu^-\, p \, \pi^+ 
\label{reac}
\eeq 
is studied. 
We adopt standard kinematics with the definitions
\[
q=k-k', \qquad Q^2=-q^2, \qquad W^2=p'{}^{2}
\]
and compute the cross section $\frac{d\si}{d Q^2 dW}$.  The mass of the resonance is not restricted to a specific value but allowed to vary within an interval proportional to the width. Consequently we let $W$ to vary and write the cross section with formulas analogous to deep inelastic scattering. The cross section is now written as 
\beq
\frac{d\si}{d\Omega dE'}=\frac{G^2}{16\pi^2}\cos^2\theta_C\frac{E'}{E}L_{\mu\nu}{\cal W}^{\mu\nu}
\eeq
with $m_N$ the mass of the nucleon in the target, $M_R$ the mass of the resonance and the leptonic tensor 

\begin{eqnarray} \di
L_{\mu\nu}=Tr[\gamma_\mu(1-\gamma_5)\slash{k}\gamma_\nu \slash{k'}] 
\\  \di
=4(k_\mu k'_\nu+k_\nu k'_\mu-g_{\mu\nu} k\cdot k' 
-i\ve_{\mu\nu\alpha\beta} k^\alpha k'{}^{\beta}) \nonumber
\end{eqnarray}

The hadronic tensor is defined as 
\beq
\begin{array}{l} \di
{\cal W}^{\mu\nu}
=\frac{1}{2m_N}\sum  \langle p | J^\mu(0)|\Delta\rangle \langle \Delta | J^\nu(0)| p \rangle \delta(W^2-M_R^2)
 \\[2mm]  \di
=-{\cal W}_1 g^{\mu\nu} + \frac{{\cal W}_2}{m_N^2} p^\mu p^\nu 
-i\ve^{\mu\nu\si\la}p_\si q_\la \frac{{\cal W}_3}{2m_N^2}
\\[2mm]  \di  \nonumber
+\frac{{\cal W}_4}{m_N^2} q^\mu q^\nu  
+\frac{{\cal W}_5}{m_N^2} (p^\mu q^\nu +q^\mu p^\nu)
+i\frac{{\cal W}_6}{m_N^2} (p^\mu q^\nu - q^\mu p^\nu)
\label{calW}
\end{array}
\eeq
where the sum implies a sum over the $\Delta$ polarization states and an averaging over the spins of the target. The integration over phase space of the $\Delta$ was carried out and gives the one-dimentional $\delta-$function. 
Sometimes it is convenient to use other variables for  resonance production
\beq
\frac{d\si}{dQ^2 dW}=\frac{\pi W}{m_N E E'} \frac{d\si}{d\Omega d E'}
\label{cs2}
\eeq
Since the $\Delta$ resonance has an observable width, the $\delta-$function should be replaced by its resonance representation 
\beq   \di
\delta(W^2-M_R^2)=\frac{M_R \Ga_R}{\pi} \frac1{(W^2-M_R^2)^2 + M_R^2 \Ga_R^2}.
\eeq
It is known that resonance production dominates neutrino reactions in the few GeV energy region. The formalism we present in this section is general and holds for various resonances. Later on, when we relate the structure functions to the form factors, we specialize to distinct final states.

The hadronic matrix element differs from resonance to resonance and contains vector and axial form factors. A convenient parametrization for the $\Delta^{++}$ resonance is the following
\beq
\langle  \Delta^{++}| J^\nu |  p \rangle  =\sqrt{3} \bar \psi_\la(p') d^{\la \nu} u(p) \quad 
\mbox{with}
\eeq
\begin{widetext}
\[
\begin{array}{l} \di
d^{\la\nu}=g^{\la\nu}\left[ \frac{C_3^V}{m_N} \slash{q}  + \frac{C_4^V}{m_N^2}(p'q) 
                 + \frac{C_5^V}{m_N^2} (pq)  + C_6^V \right]\ga_5
-q^\la \left[ \frac{C_3^V}{m_N} \ga^\nu  + \frac{C_4^V}{m_N^2} p'{}^\nu 
            + \frac{C_5^V}{m_N^2} p^\nu  \right] \ga_5
\\[6mm] \hspace*{12mm} \di
+ g^{\la\nu}\left[ \frac{C_3^A}{m_N} \slash{q} + \frac{C_4^A}{m_N^2} (p'q)  \right]
- q^\la \left[ \frac{C_3^A}{m_N} \ga^\nu  + \frac{C_4^A}{m_N^2} p'{}^\nu  \right]
+ g^{\la\nu} C_5^A  + q^\la q^\nu \frac{C_6^A}{m_N^2}. 
\end{array}
\]

In the square of the matrix element also appears the Rarita-Schwinger projection operator

\beq \di
\left| \psi_\Delta \right\rangle \left\langle \psi_\Delta \right| =
S^{\si\la} = 
[\slash{\W}+M_R]
\left(
-g^{\si\la}  +\frac13 \ga^\si \ga^\la  +\frac1{3M_R}(\ga^\si \W^\la - \W^\si \ga^\la )
              + \frac{2}{3M_R^2} \W^\si \W^\la 
\right).
\label{propagator-32}
\eeq
\end{widetext}

With these preliminaries the hadronic tensor takes the form 
\beq
{\cal W}^{\mu\nu}
=\frac32 \frac{1}{2m_N} Tr\left[ (\bar{d})^{\mu\si} S_{\si \la} d^{\la \nu} (\slash{p}+m_N) \right] \delta(W^2-M_R^2)
\label{wfordelta}
\eeq
with $(\bar{d})^{\mu\si} = \gamma_0 (d^+)^{\mu\si} \gamma_0$ and then parametrized according to (\ref{calW}). This way we define the relative normalization between the structure functions and the form factors.
The factor $3$ comes from the isospin coefficient for $\Delta$ and the $1/2$ from the averaging over the initial spins of the target. The results of the calculation are summarized in Appendix~A.

The remaining problem consists in writing the cross section in terms of form factors and specifying their numerical strength and $Q^2-$dependence. The cross section assumes the standard form which includes now the mass of the muon. 

\begin{widetext}
\begin{eqnarray} \di
\frac{d\si}{dQ^2 dW}=\frac{G^2}{4\pi}\cos^2\theta_C\frac{W}{m_N E^2} \Biggl\{
{\cal W}_1(Q^2+m_\mu^2)
+\frac{{\cal W}_2}{m_N^2}\left[ 2(\pk)(\pkprime) - \frac12 m_N^2 (Q^2+m_\mu^2)\right]
\nonumber \\
-\frac{{\cal W}_3}{m_N^2} \left[ Q^2\pk - \frac12\pq (Q^2+m_\mu^2) \right]
+\frac{{\cal W}_4}{m_N^2}m_\mu^2 \frac{(Q^2+m_\mu^2)}{2}
-2\frac{{\cal W}_5}{m_N^2} m_\mu^2 (\pk) \Biggr\}
\label{cross-sec}
\end{eqnarray}
\end{widetext}
The dependence on the muon mass agrees with the one in Ref.  \cite{Albright:1974ts}.
The structure functions are also expressed in terms of the form factors. This is straightforward and for $P_{33}(1232)$ resonance leads to the relations given in Appendix. For final states with opposite parity, like $D_{13}(1520)$, the $\gamma_5$ matrices in the current-nucleon--resonance vertex will appear as multiplicative factors to the axial (and not the vector) form factors. The effect of this change to Eqs.(\ref{calW1})--(\ref{calW6}) is the replacement of $m_N M_R$ by $-m_N M_R$.

The determination of the form factors follows from general principles and experimental results. We begin with the vector form factors. The conserved vector current hypothesis gives the relation $C_6^V=0$. The remaining form factors also occur in electroproduction where it has been established that the $M_{1+}$ multipole dominates. Recent data determine the contribution from the electric multipole $E_2$ to be $\sim -2.5\%$ and from the scalar multipole $\sim -5\%$ \cite{Joo:2001tw}. We shall assume the dominance of the magnetic dipole which gives 
\[
C_3^V=1.95, \qquad C_4^V=-C_3^V\frac{m_N}{W}, \qquad C_5^V=0.
\]
The  numerical value is obtained from the data in electroproduction after an isospin rotation. Electroproduction data lead to a $Q^2-$dependence faster than the dipole \cite{Paschos:2003qr}
\[
C_3^V(Q^2)=\frac{C_3^V(0)}{\left(1+ Q^2 / M_V^2\right)^2}\, \frac1{1+ Q^2 / 4M_V^2}, 
\]
with $M_V=0.84\GeV$. This functional dependence indicates that the size of the resonance is larger because of the mesonic cloud surrounding the resonance and its  Fourier transform gives a steeper function of $Q^2$.

Among the axial form factors the most important contribution comes from $C_5^A$ whose numerical value is related to the pseudoscalar form factor $C_6^A$ by PCAC. We shall use the values 
\beq
\begin{array}{c} \di
C_5^A(0)=\frac{f_\pi g_\Delta}{\sqrt{3}} = 1.2, \quad C_4^A=-\frac{C_5^A}{4}, 
\\ \di C_3^A=0, \quad C_6^A=C_5^A\frac{m_N^2}{Q^2+m_\pi^2} 
\end{array}
\label{ff-axial}
\eeq
with  $g_\Delta=15.3$, $f_\pi=0.97 m_\pi$
and 
\beq
C_5^A(Q^2)=\frac{C_5^A(0)}{\left(1+ Q^2 / M_A^2\right)^2}\, \frac1{1+ Q^2 / 3M_A^2},
\label{axial-modif-dipole}
\eeq
The value for $C_4^A$ was found to give a small contribution to the cross section and the dipole form factor for $C_5^A$ is again modified. The form factor $C_3^A$ is set to zero as suggested by early \cite{Salin:1967aa,Adler:1968tw,Bijtebier:1970ku} dispersion calculations.
It remains to introduce the functional form for the width of the resonance
\beq
\Gamma=\Gamma_0 \left( \frac{p_\pi(W)}{p_\pi(M_R)} \right)^{(2l+1)}, 
\label{gamma2l}
\eeq
with $l=1$ for $P_{33}(1232)$ resonance. A partial width of the form 
\beq
\Gamma=\Gamma_0 \frac{p_\pi(W)}{p_\pi(M_R)}.
\label{gamma1}
\eeq
was also used \cite{Schreiner:1973mj,Paschos:2003qr}, but now Eq.(\ref{gamma2l}) is preferable since it is required by the partial wave analysis.

With these results and those in Appendix A, one has a complete set of formulas with which to proceed to analyse the electromagnetic and weak production of the $P_{33}$ resonance.

\section{Experimental results\label{experresults}}

A detailed experimental study, including the $Q^2-$dependence of the differential neutrino cross section, was made in two experiments: using the Argonne National Laboratory 12-ft bubble chamber (ANL)\cite{Radecky:1981fn} and the Brookhaven National Laboratory 7-ft bubble chamber (BNL) \cite{Kitagaki:1986ct,Kitagaki:1990vs}. In both experiments the neutrino spectrum was peaked at approximately $1\GeV$. We calculate the cross section for the reaction $\nu \, p \to \mu^-\, \Delta^{++}  \to \mu^-\, p \, \pi^+ $ weighted over the neutrino spectrum. 

\begin{figure}[t]
\begin{center}
\includegraphics[angle=-90,width=\columnwidth]{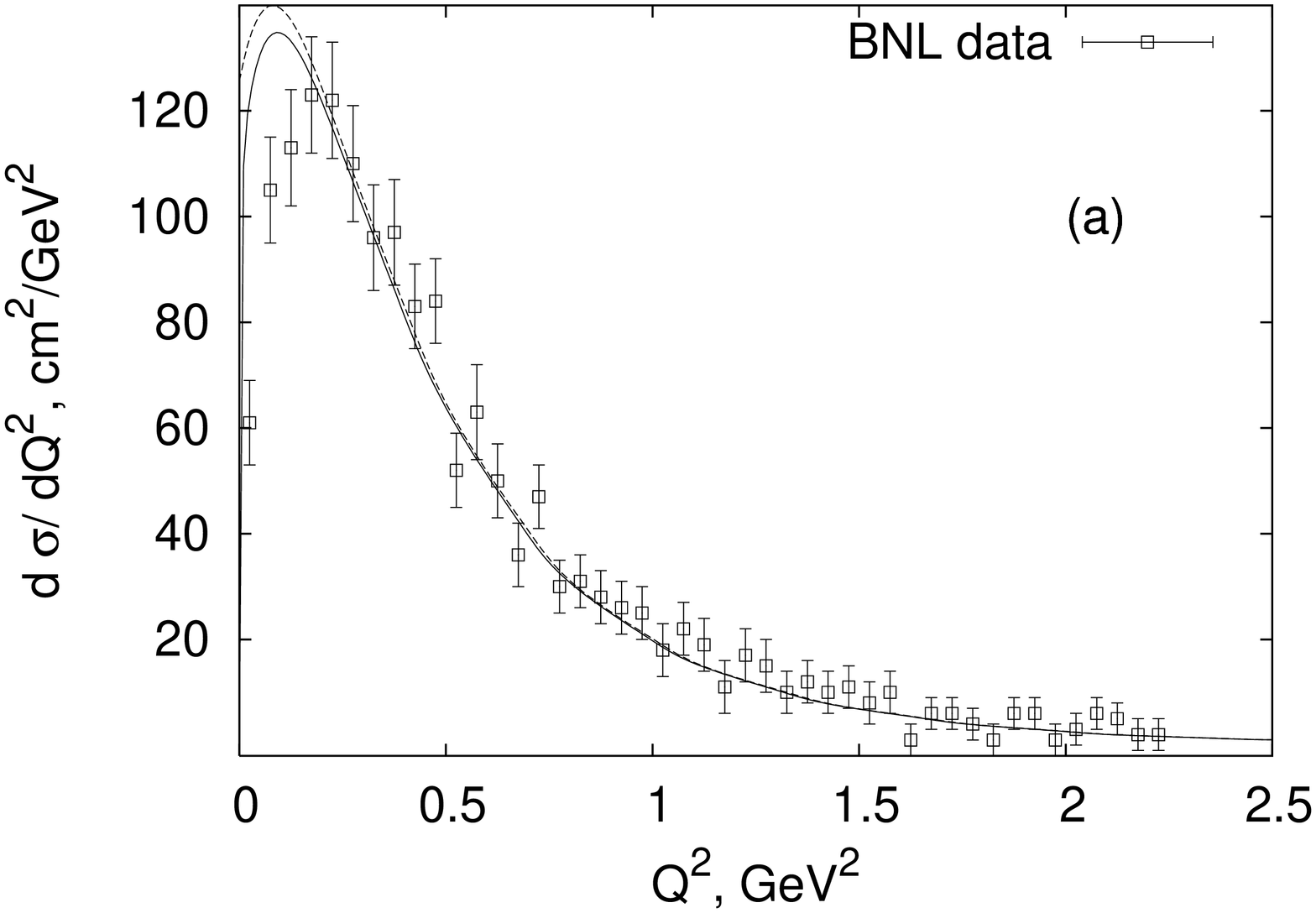} $\; \;$
\includegraphics[angle=-90,width=\columnwidth]{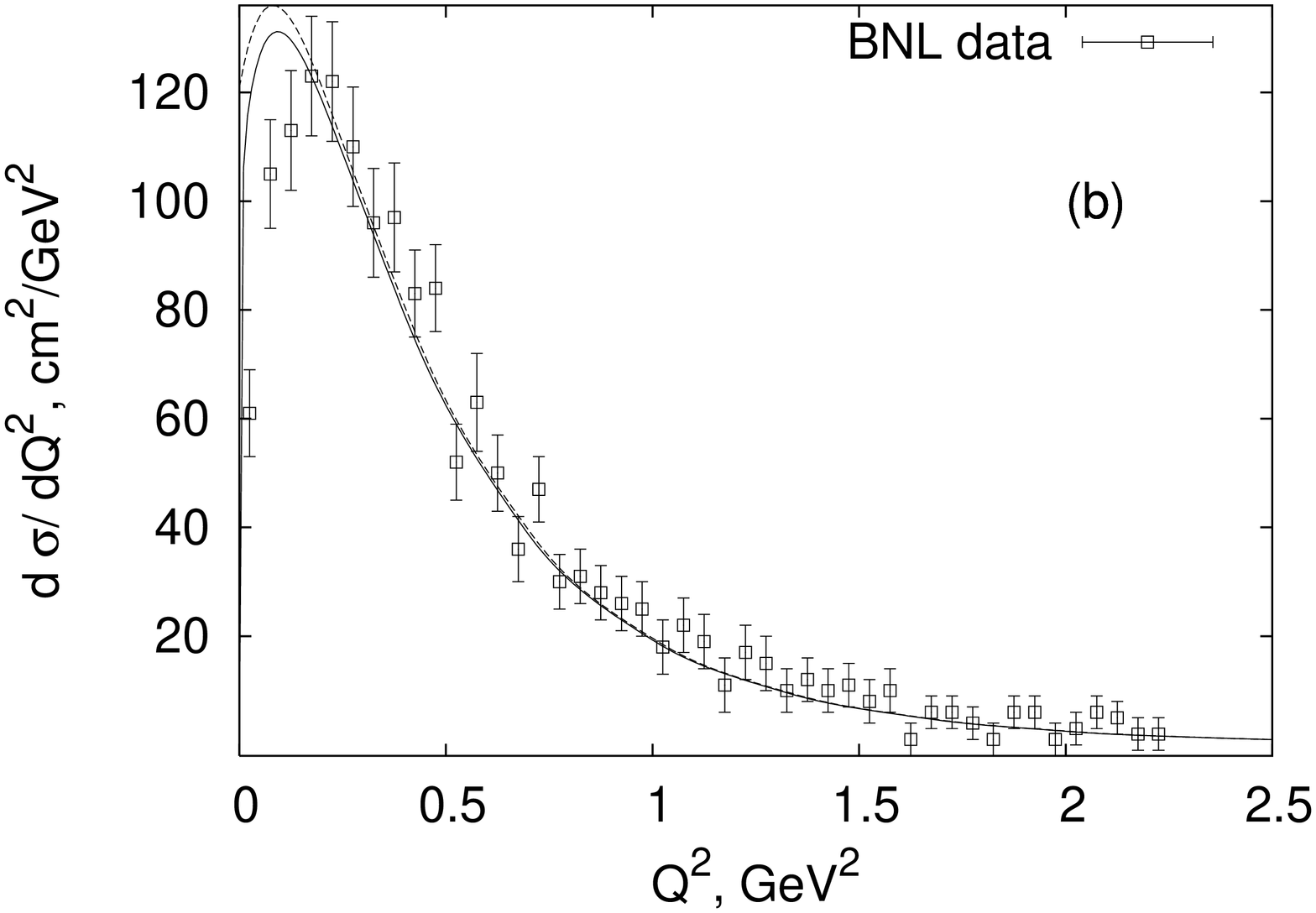}
\end{center}
\caption{The cross section $\di {d \si}/{d Q^2}$, calculated for the BNL neutrino energy spectrum and compared with the experiment for the running width (\ref{gamma2l}) (Fig.~(a)) and (\ref{gamma1}) (Fig.~(b)). The full lines are for  the case $m_\mu=0.105\GeV$, the dashed lines are for the approximation $m_\mu=0$.}
\label{BNL-Q2}
\end{figure}

The BNL experiment observed  a peak in the differential cross section $d\si/ dQ^2$  at about $Q^2=0.175\GeV^2$ as shown in Fig.~\ref{BNL-Q2}, where experimental data are presented in arbitrary units.  We computed the $Q^2-$distribution with the form factors described above and $M_A=1.05\GeV$. The results are  shown  in Fig.~\ref{BNL-Q2}. The theoretical curve $d\si/ dQ^2$ has a peak at $Q^2=0.085 \GeV^2$.  

It is evident that the agreement between theory and experiment at $Q^2>0.2\GeV^2$ is satisfactory for both functional forms of the resonance width (\ref{gamma2l}) and (\ref{gamma1}). For the overall scale, we normalize the area under the theoretical curve for $Q^2>0.2\GeV^2$ to the corresponding curve under the data. It is also evident that the muon mass brings an additional  decrease in the region of small $Q^2$ where the Pauli  suppression is also significant, but the data are still slightly lower than the theoretical curve.

\begin{figure}[b]
\begin{center}
\includegraphics[angle=-90,width=\columnwidth]{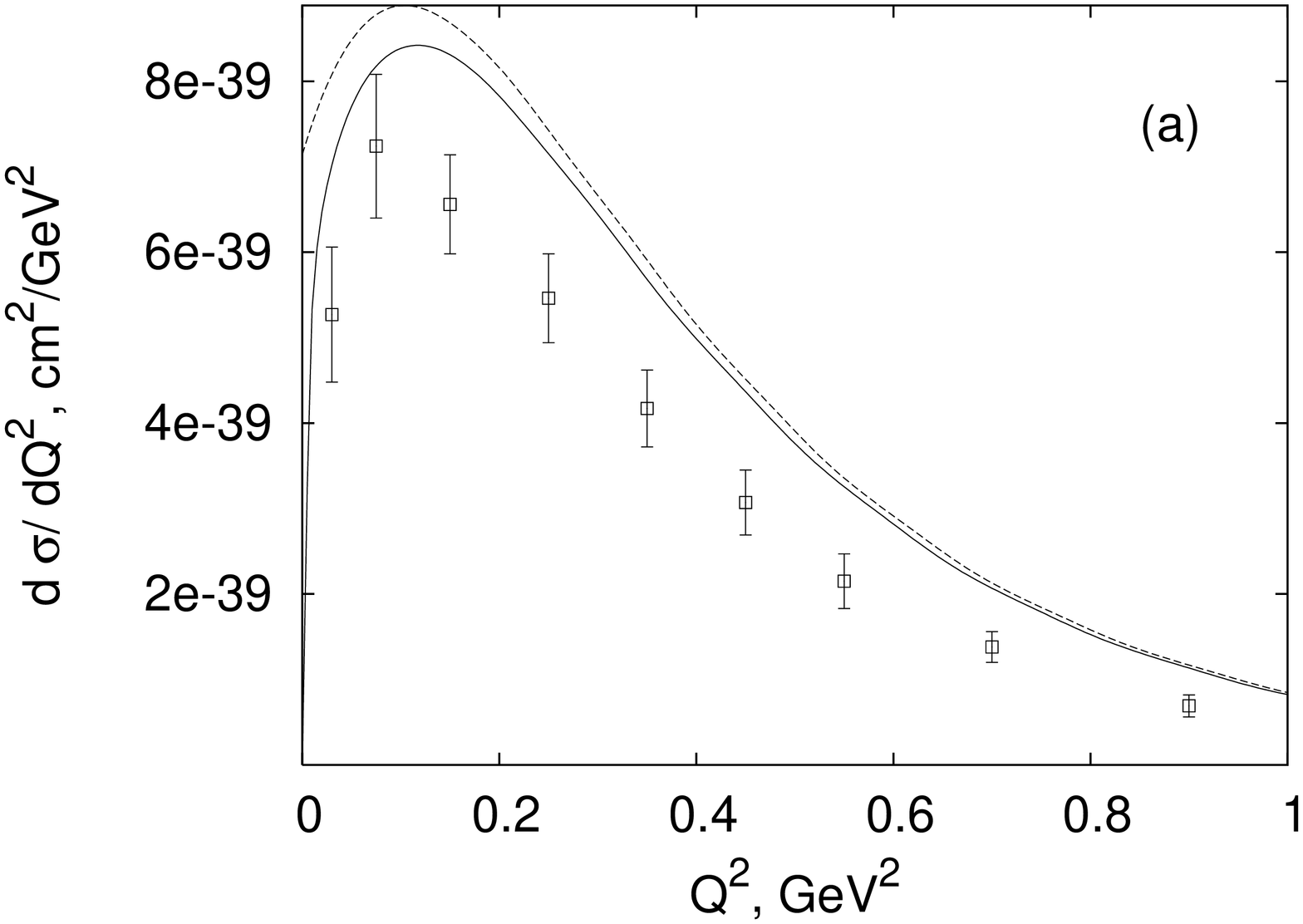} 
\includegraphics[angle=-90,width=\columnwidth]{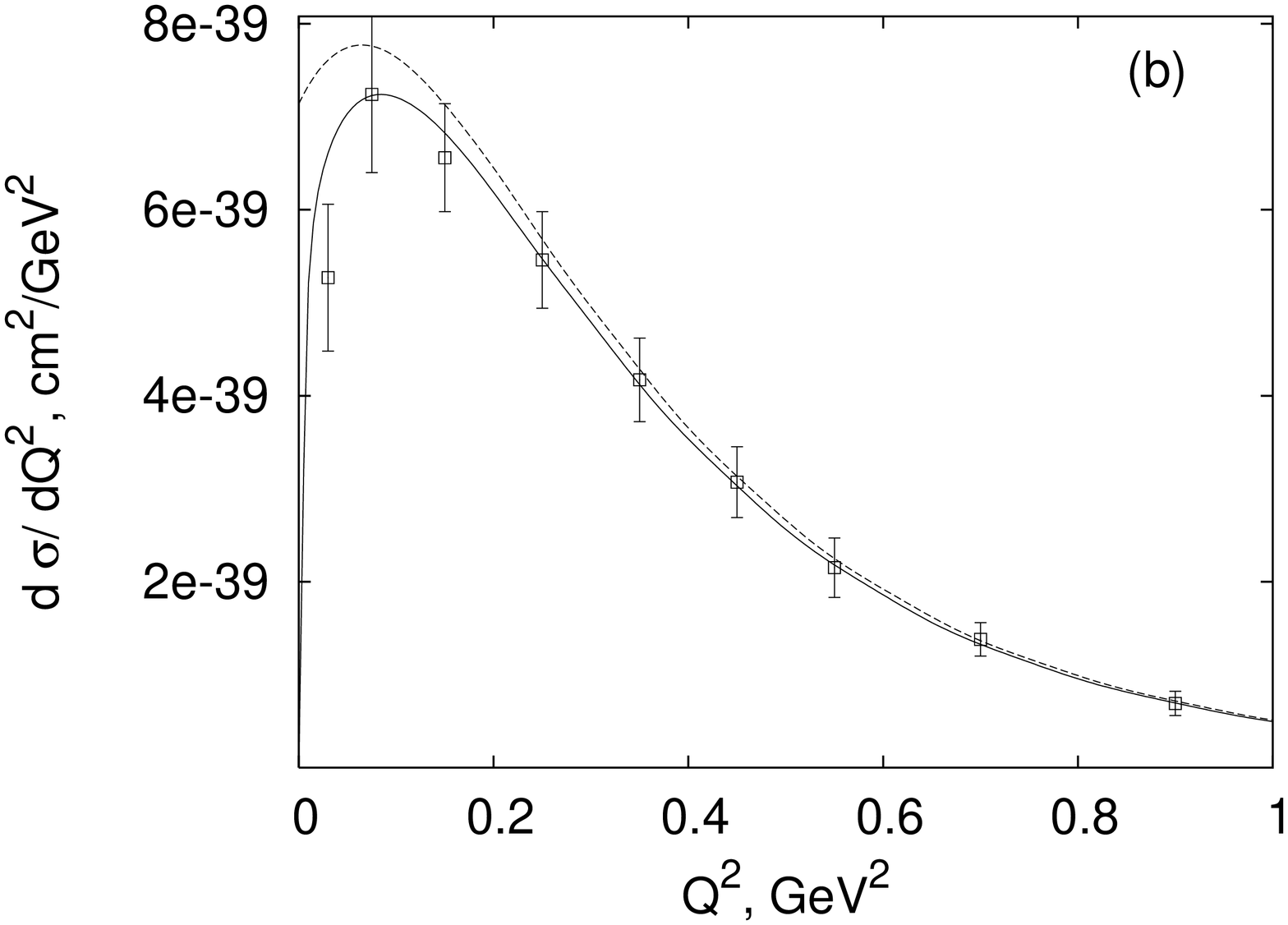}
\end{center}
\caption{The cross section $\di d \si / d Q^2$, calculated for the ANL neutrino energy distribution.
The full lines are for $m_\mu=0.105\GeV$, the dashed lines are for the approximation $m_\mu=0$.}
\label{ANL-Q2}
\end{figure}

In the ANL experiment the data are  with large bins of $Q^2$ and the maximum of $d\si/ dQ^2$ is at a larger value of $Q^2$. The formalism described so far determines the cross section including the absolute normalization. For $M_A=1.05\GeV$ and the modified dipole in Eq.(\ref{axial-modif-dipole}) we obtain the curve in Fig.~\ref{ANL-Q2}a, which is above the data. The integrated cross section in this case at high energies approaches $0.7\cdot 10^{-38}\cm^2$, which is consistent with the experimental data.  The discrepancy in $Q^2-$dependence can not be resolved by the overall normalization of the curve and requires a decrease of $M_A=0.84\GeV$ in order to obtain the curve in Fig.~\ref{ANL-Q2}b.  The two curves are without (dotted) and with (solid curve) the muon mass. The integrated cross section is also decreased approaching at high energies a constant value of $0.55\cdot 10^{-38}\cm^2$, which is also consistent with the data.  

An earlier theoretical analysis \cite{Alvarez-Ruso:1998hi} accounts for the ANL data by using similar couplings and muon mass effects. They include nuclear corrections by using deuterium wave functions and compare the differential cross section to the ANL data. 
Another approach \cite{Amaro:2004bs} describes electron
and neutrino scattering  on various nuclei in  terms of a scaling law abstracted 
from data and the authors present several distributions.  A direct comparison with 
our results is not available and perhaps difficult because of the different  methods.

Another way to reach an agreement with the data is to replace the dependence (\ref{axial-modif-dipole}) with a steeper dependence, for example
\[
C_5^A(Q^2)=\frac{C_5^A(0)}{\left(1+ Q^2 / M_A^2\right)^2}\, \frac1{1+ 2Q^2 / M_A^2}, 
\]
with $M_A=1.05\GeV$
or 
\[
C_5^A(Q^2)=\frac{C_5^A(0)}{\left(1+ Q^2 / M_A^2\right)^2}\, \left(\frac1{1+ Q^2 / 3M_A^2}\right)^2,
\]
with $M_A=0.95\GeV$

The theoretical formalism is very close to explaining the experimental data. There are differences in the BNL and ANL data which we can not understand and which must be resolved by future experiments. In addition there is  a sharp decrease at small $Q^2<0.2\GeV^2$ which appears to be persistent. One part  of the decrease comes from the Pauli suppression, which is small for deuterium and another part from terms depending on the muon mass which are important for the low energy of the neutrino beam.

One can easily see from Fig.~\ref{ANL-Q2}, that for low energies, i.e. neutrino energy $E\sim 1\GeV$, taking into account the nonzero muon mass reduces the cross section at small $Q^2$ by approximately $20\%$. The physical origin of this reduction is as follows. Firstly, the double differential cross section  $d\si / dQ^2 dW$ changes mainly due to the contribution from ${\cal W}_5$ and ${\cal W}_4$ structure functions.  Secondly, for each $Q^2$ we must integrate this cross section over $W$. The lower limit of integration $W_{-}(Q^2)=m_N+m_\pi$ is independent of $Q^2$ and the muon mass. The upper limit  of integration, however, depends on $Q^2$ and  the muon mass and is given by
\[
\begin{array}{r} \di 
W_{+}^2(Q^2)=\Biggl[ \frac14 s^2 a_-^2\left(\frac{m_\mu^4}{s^2}-2\frac{m_\mu^2}{s}\right) 
- \left(Q^2+\frac12 m_\mu^2 a_+^2 \right)^2 
\\  \di
+ s \, a_{-} \left( Q^2 + \frac{m_\mu^2}{2} a_+ \right) \Biggr] \left/ \left[a_- (Q^2+m_\mu^2)\right], \right.
\end{array}
\]
where $s=2m_N E +m_N^2$, $a_{\pm}=1 \pm m_N^2/s$.

\begin{figure}[h]
\begin{center}
\includegraphics[angle=-90,width=\columnwidth]{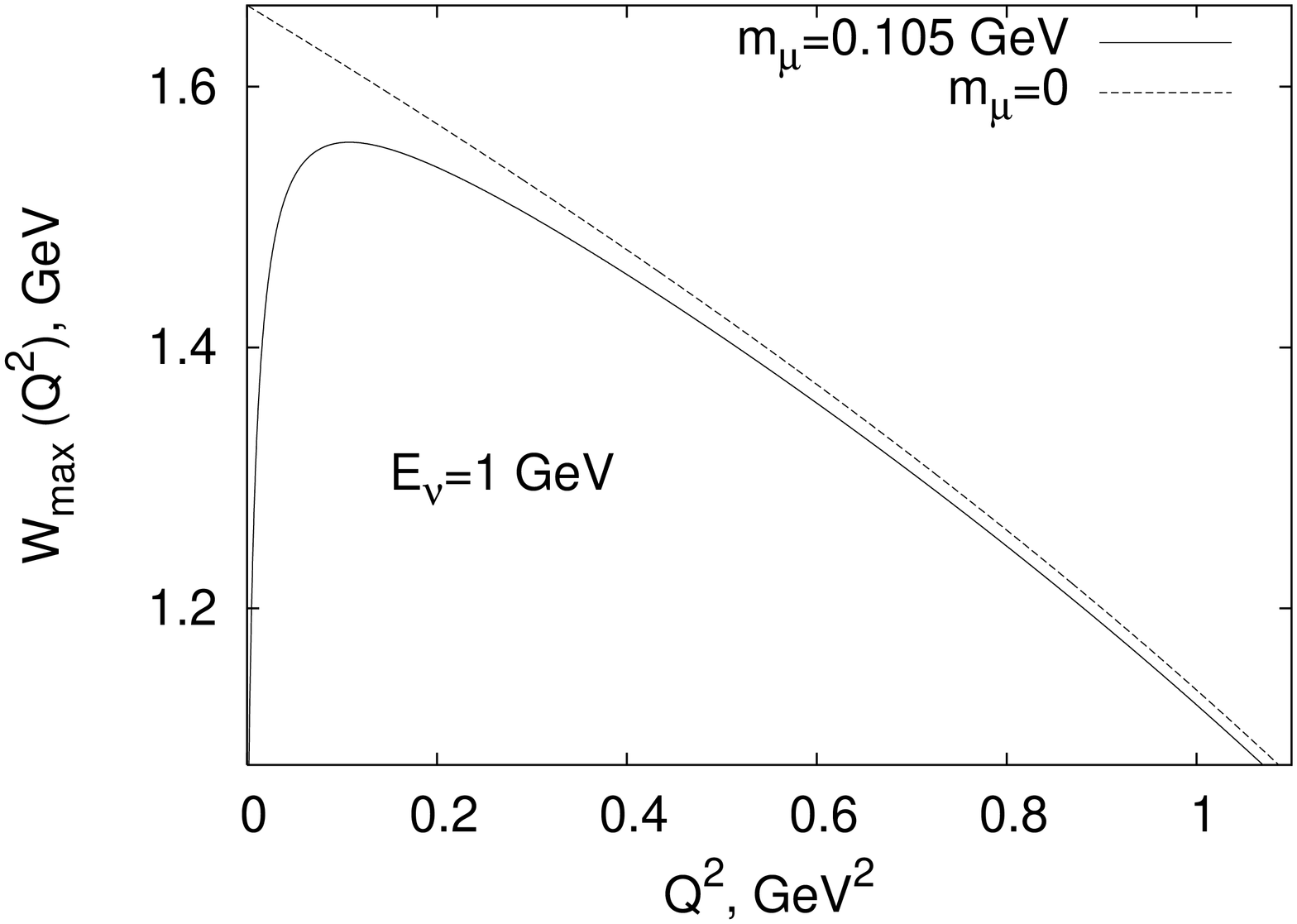} $\; \;$
\includegraphics[angle=-90,width=\columnwidth]{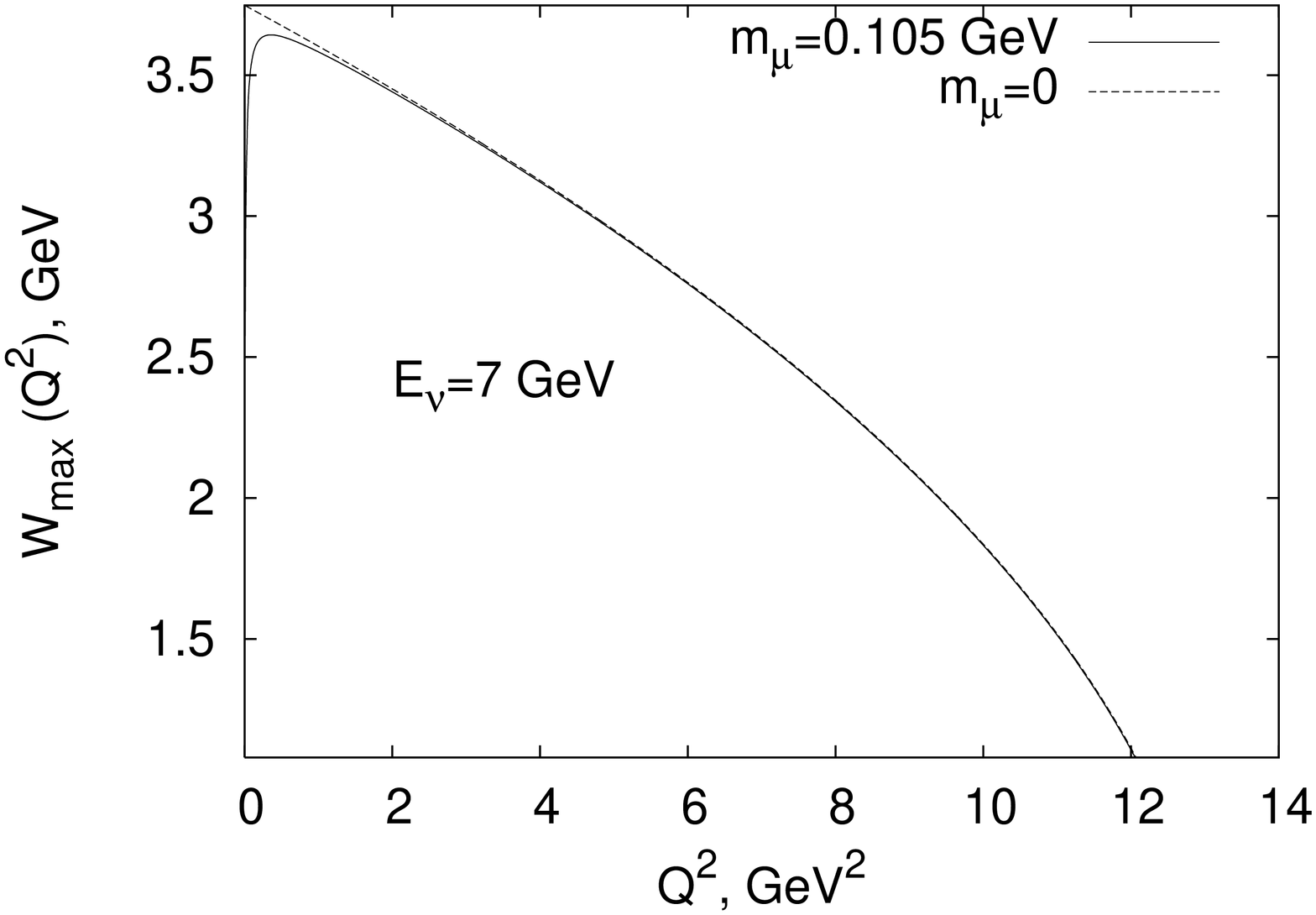}
\end{center}
\caption{Upper integration limit $W_{+}(Q^2)$ for $E=1\GeV$  and $E=7\GeV$  The full lines are for $m_\mu=0.105\GeV$, the dashed lines are for the approximation $m_\mu=0$.}
\label{intlimits}
\end{figure}

The integration limits are shown in Fig.~\ref{intlimits} for $E=1\GeV$  and $E=7\GeV$. One could easily notice that taking into account the muon mass noticeably decreases $W_{+}$ and implies the reduction of $d\si / dQ^2$. This effect diminishes as the neutrino energy increases. In the rest of the section the muon mass is taken into account, but not discussed any more.

\begin{figure}[hbt]
\includegraphics[angle=-90,width=\columnwidth]{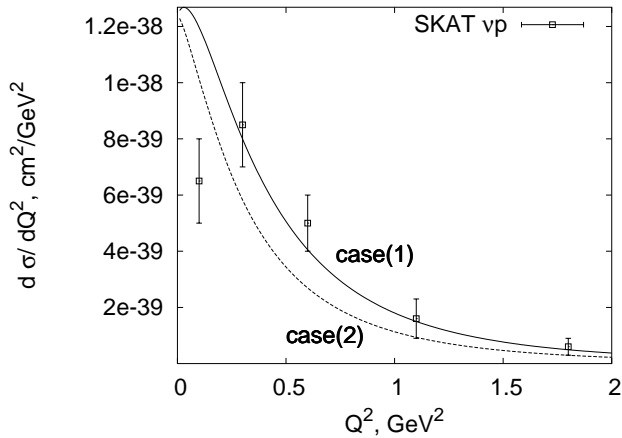} 
\caption{The cross section $\di d \si / d Q2$, calculated for the SKAT experiment with $\langle E \rangle = 7\GeV$ for the behavior of the form factors in cases (1) and (2).}
\label{skat}
\end{figure}

For the sake of completeness we mention experiments at high energies. In these cases  effects from the muon mass are diminished so that the levelling of the $d\si / dQ^2$ distribution should disappear. 
The SKAT \cite{Ammosov:1988xb} experiment had an average neutrino energy of $E=7\GeV$. The results are shown in Fig.~\ref{skat} together with theoretical curves with $M_A=1.05\GeV$ (which we call case (1)) and with $M_A=0.84\GeV$ (case (2)). There are few experimental points and the error bars are too large to draw conclusions.

\begin{figure}[hbt]
\includegraphics[angle=-90,width=\columnwidth]{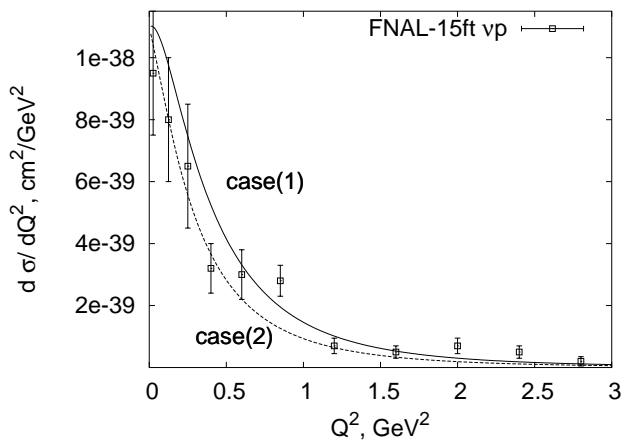}
\caption{The cross section $\di d \si / d Q2$, calculated for the FNAL experiment for the behavior of the form factors in cases (1) and (2).}
\label{fnal}
\end{figure}

In the FNAL 15-ft bubble chamber \cite{Bell:1978rb} experiment data are available for neutrino energies between $15$ and $40\GeV$. At such energies the integrated cross section remains constant with high accuracy, so the exact value of neutrino energy is  not important. The data and the theoretical curves are presented in Fig.~\ref{fnal}. 

\begin{figure}[hbt]
\begin{center}
\includegraphics[angle=-90,width=\columnwidth]{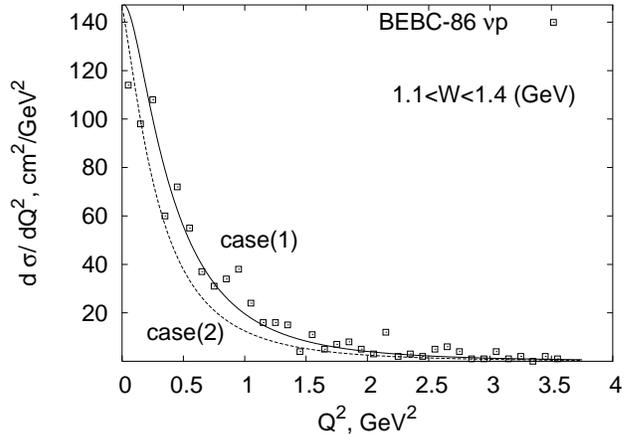} 
\includegraphics[angle=-90,width=\columnwidth]{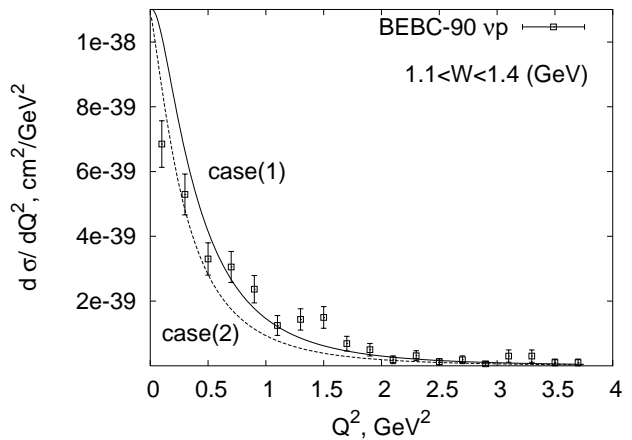}
\end{center}
\caption{The cross section $\di {d \si}/{d Q2}$ in experiments BEBC-86 and BEBC-90 for the behavior of the form factors in cases (1) and (2).}
\label{bebc}
\end{figure}

From the BEBC experiment, obtained with the CERN wide-band beam, two data sets are available:  BEBC-86 \cite{Allen:1985ti} with $\langle E \rangle = 24.8\GeV$  and BEBC-90 \cite{Allasia:1990uy} (the neutrino flux is given in \cite{Allasia:1983dq}) with $\langle E \rangle = 54\GeV$. They are shown in Fig.~\ref{bebc}. A common property is the disappearance of the flattening of the cross section at small $Q^2$, as expected.
 
We conclude, that different experiments, performed with the help of bubble chambers in the 80's, show at low $Q^2$ a slightly lower cross section than theoretically predicted. The experiments described so far are not detailed enough to allow separation of the form factors and a unique determination of their $Q^2$ dependence.

\begin{figure}[hbt]
\includegraphics[angle=-90,width=\columnwidth]{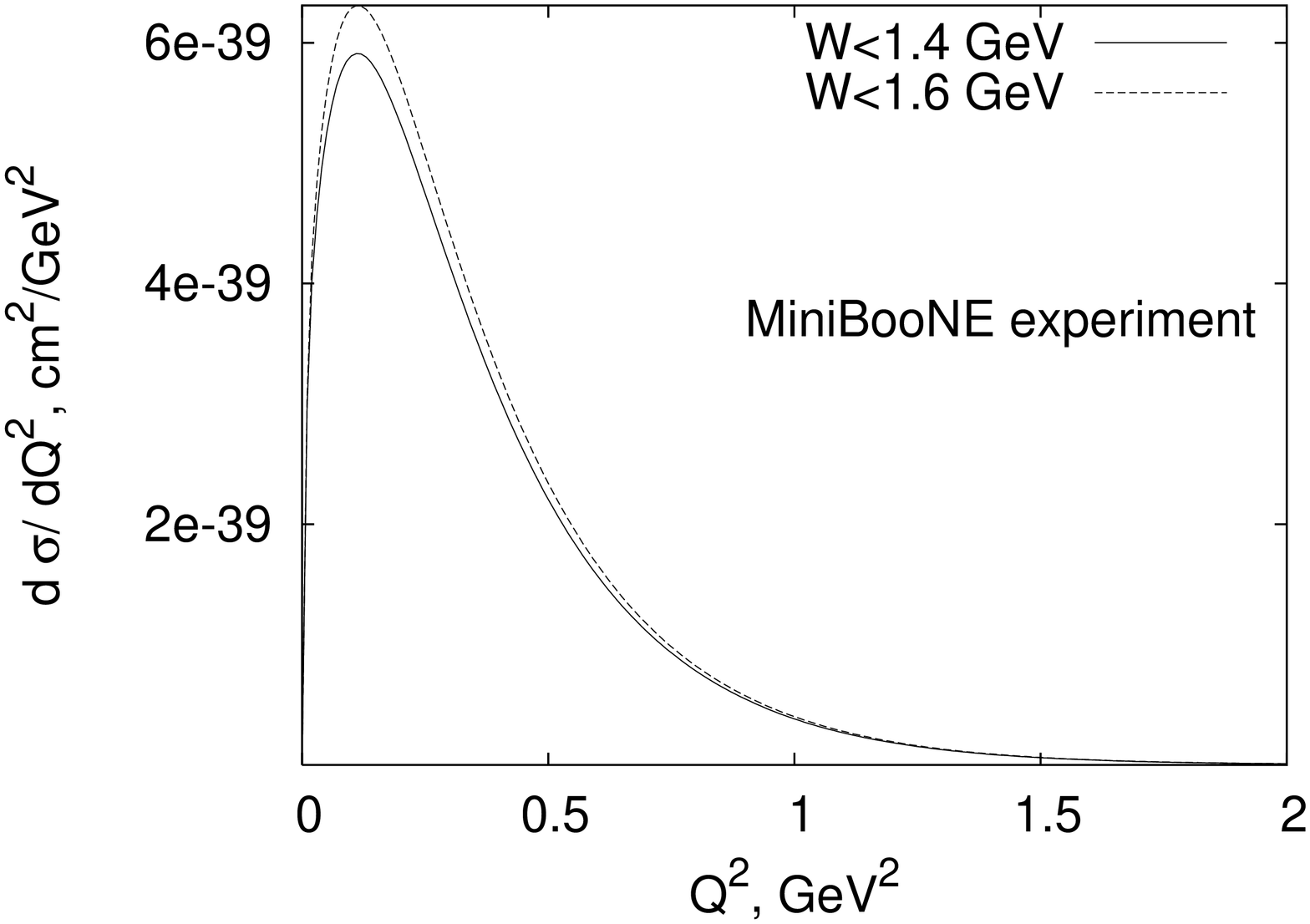} 
\caption{The cross section $\di d \si / d Q2$, predicted for MiniBOONe experiment, $W<1.4$ (solid line) and $W<1.6$ (dashed line).}
\label{miniboone}
\end{figure}

\begin{figure}[htb]
\includegraphics[angle=-90,width=\columnwidth]{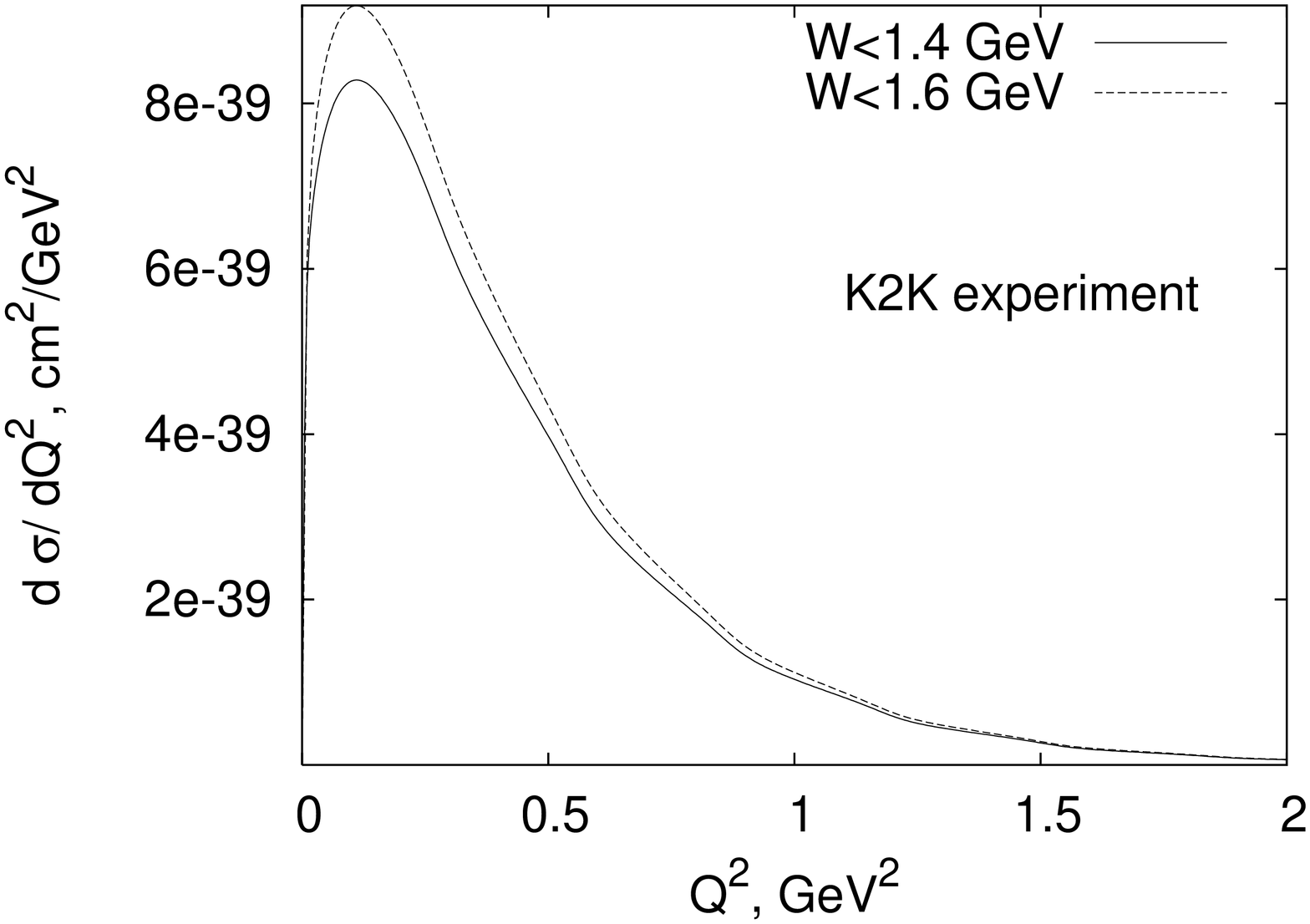}
\caption{The cross section $\di d \si / d Q2$, predicted for K2K experiment, $W<1.4$ (solid line) and $W<1.6$ (dashed line).}
\label{k2k}
\end{figure}

Two new experiments K2K and MiniBooNE will be delivering results. They are both at low neutrino energies where the muon mass effects should be important. With the neutrino spectra from \cite{Raaf:2004aa} and \cite{Gran:2004aa} we predict the $Q^2-$distributions shown in Figs.~\ref{miniboone} and \ref{k2k}, using the axial form factor in Eq.(\ref{axial-modif-dipole}).  These experiments use medium or heavy nuclei as targets and nuclear corrections must be applied, which were left out in our curves (only Pauli blocking is included).

\section{Special properties\label{specialprop}}

It is evident from our presentation that the cross section in the $\Delta$ resonance region has several important features still to be investigated. One of them deals with the structure of the form factors, especially the axial form factors. We showed that the dominant contribution comes from $C_5^A$ and $C_6^A$. Closely related is the $Q^2-$dependence of the process, especially in the small $Q^2$ region, where various effects are present:

(i) the mass of the muon is important at low energies,

(ii) nuclear corrections are also important, like the Pauli factor. Up to now there is no sign of a possible distortion of the angular distribution due to charge exchange effects, and  

(iii) effects from the mass of the muon should become more evident in the MiniBooNE and K2K experiments. 

\begin{figure}[hbt]
\begin{center}
\includegraphics[angle=-90,width=\columnwidth]{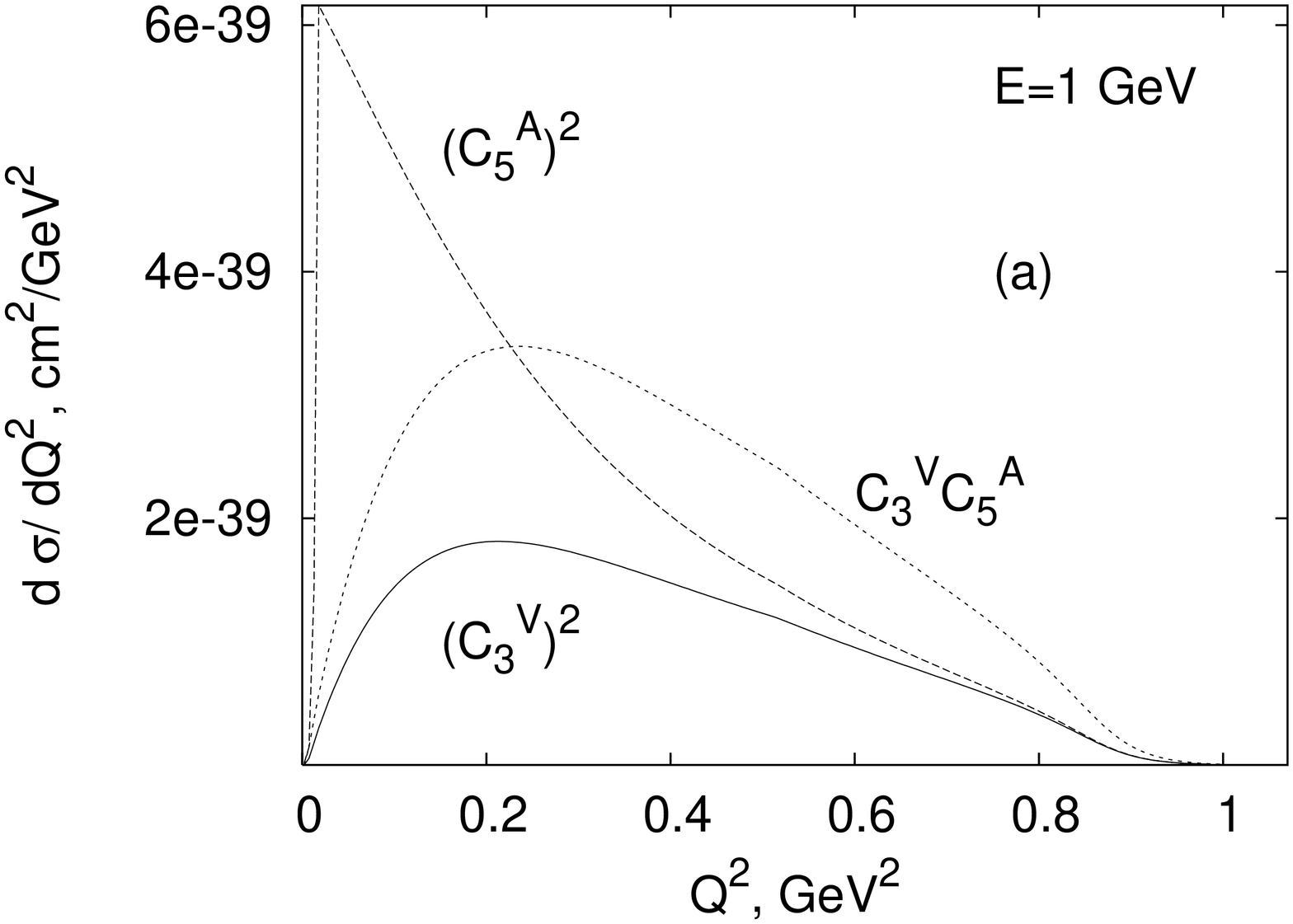} 
\includegraphics[angle=-90,width=\columnwidth]{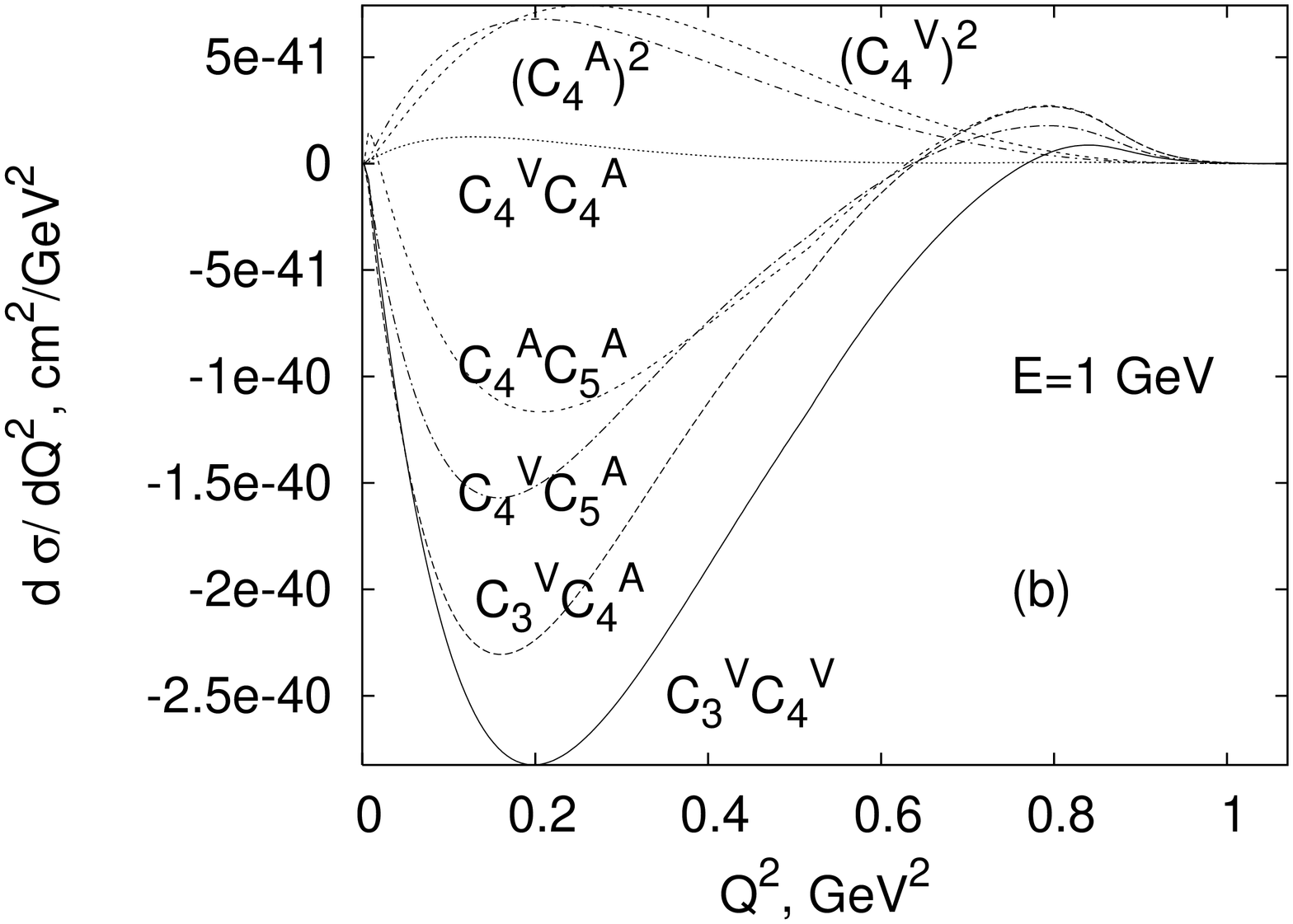}
\end{center}
\caption{Contribution of form factors to the differential cross section.}
\label{c3v5a}
\end{figure}

In order to see the relative importance of the form factors we computed in Fig.~\ref{c3v5a}a the various contributions for $E_\nu=1\GeV$.  The dominance of $C_5^A$ and $C_3^V$ is evident with the other terms contributing less than $2\cdot 10^{-40} \cm^2/\GeV^2$. As $Q^2\to 0$ only the $C_5^A$ contribution to  $d\si / dQ^2$ is dominant and remains large for very small values of $Q^2\approx 0.01 \GeV^2$ and then turns sharply to zero. This is caused by the vanishing of the phase space, as  we mentioned earlier.  Fig.~\ref{c3v5a}b  shows the contribution of the smaller form factors and the sum of them is negative with a negative valley at $Q^2\sim 0.15 \GeV^2$. 

\begin{figure}[hbt]
\begin{center}
\includegraphics[angle=-90,width=\columnwidth]{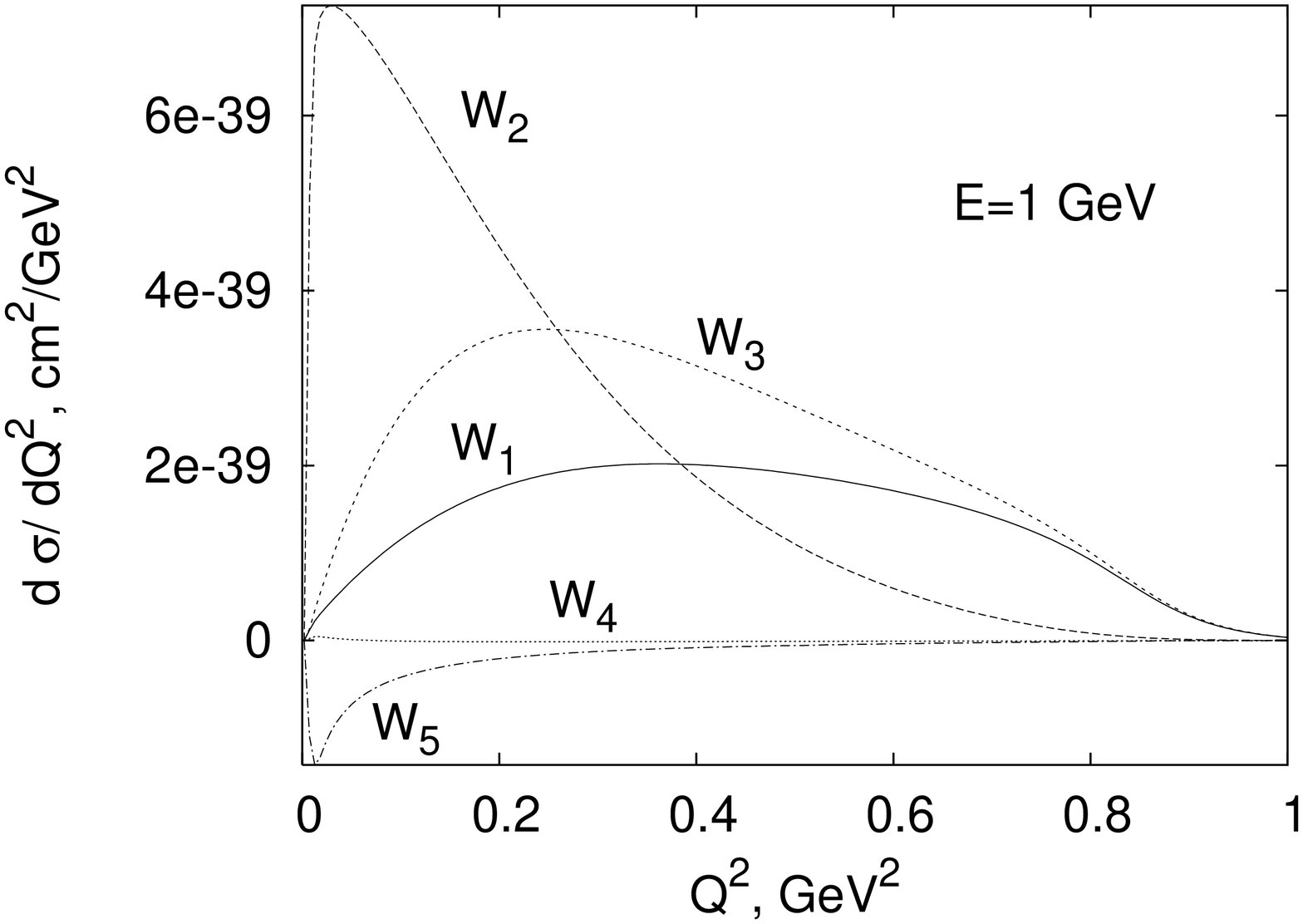} 
\includegraphics[angle=-90,width=\columnwidth]{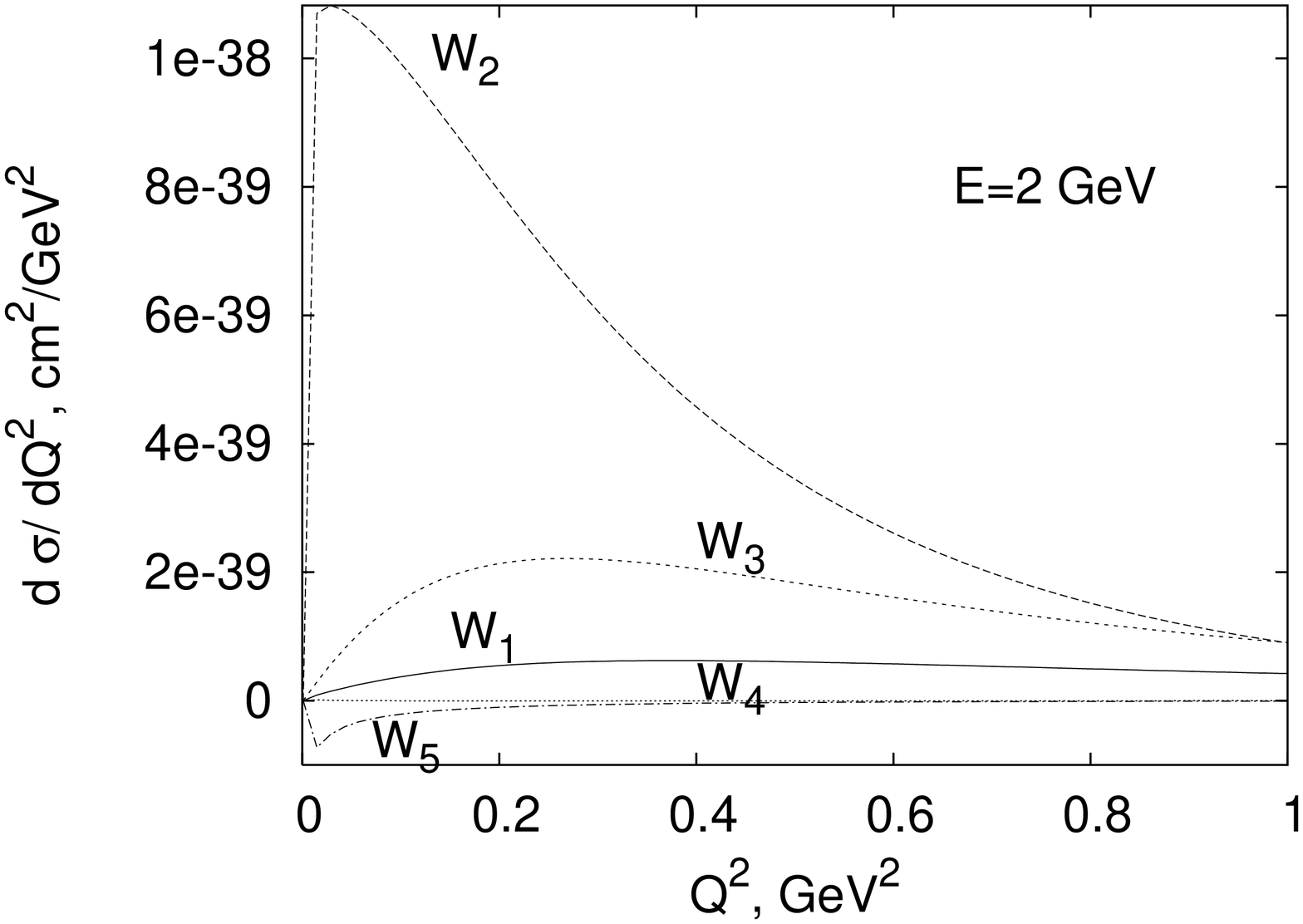}
\end{center}
\caption{Contribution of structure functions to the differential cross section.}
\label{W1W5}
\end{figure}

A similar study for the structure functions is shown in Fig.~\ref{W1W5} for two values of the neutrino energy ($E_\nu=1.0$ and $2.0 GeV$). We note that the terms from ${\cal W}_4$  and ${\cal W}_5$ are  negative and   ${\cal W}_5$ contributes to the sharp  decrease of the cross section at small values of $Q^2$.

\begin{figure}[hbt]
\begin{minipage}[c]{0.48\textwidth}
\includegraphics[angle=-90,width=\textwidth]{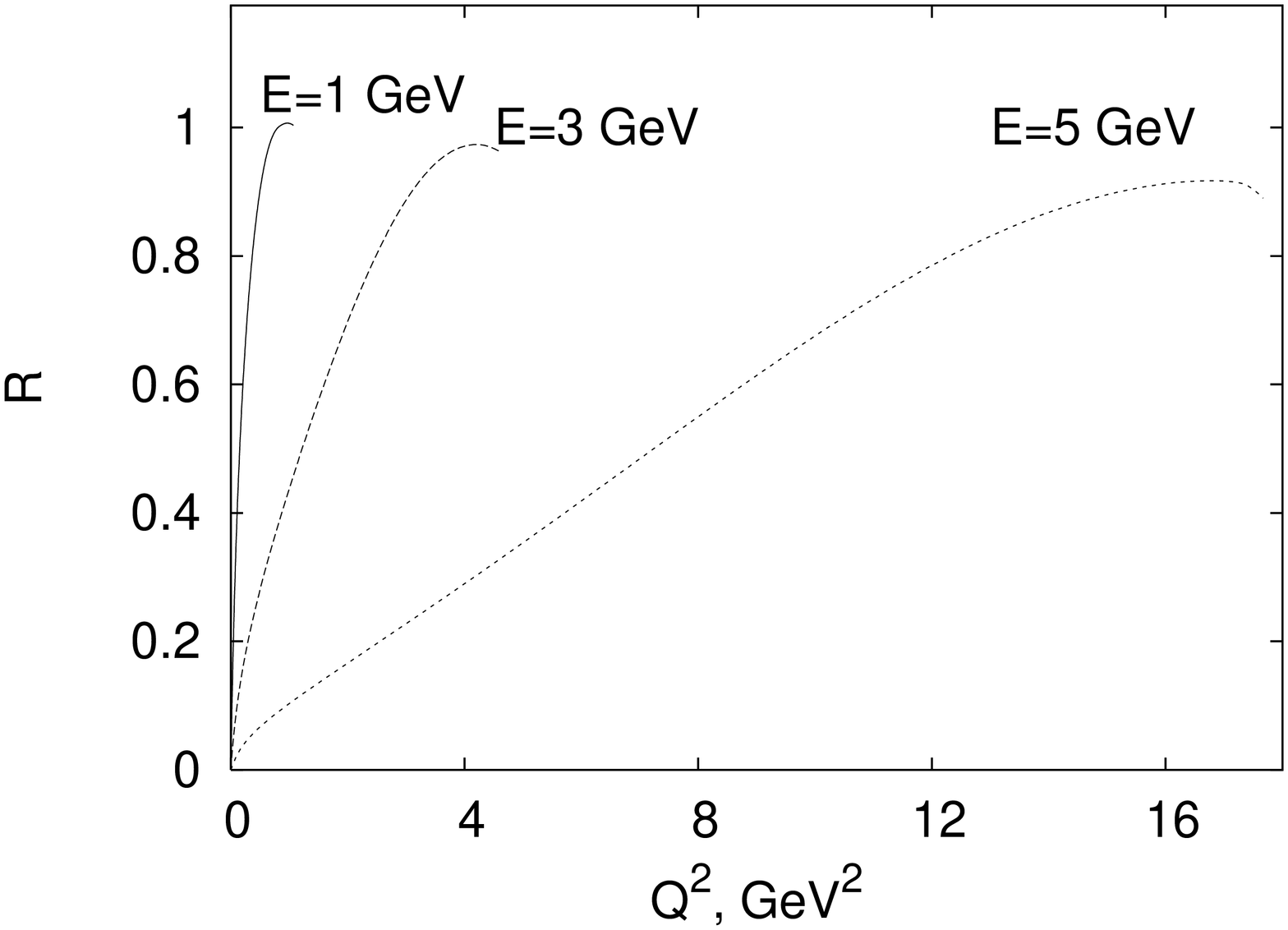} 
\caption{Predictions for $R$ for $E=1,\, 3, \, 5 \GeV$.}
\label{pred-R}
\end{minipage}
$\quad$
\begin{minipage}[c]{0.48\textwidth}
\includegraphics[angle=-90,width=\textwidth]{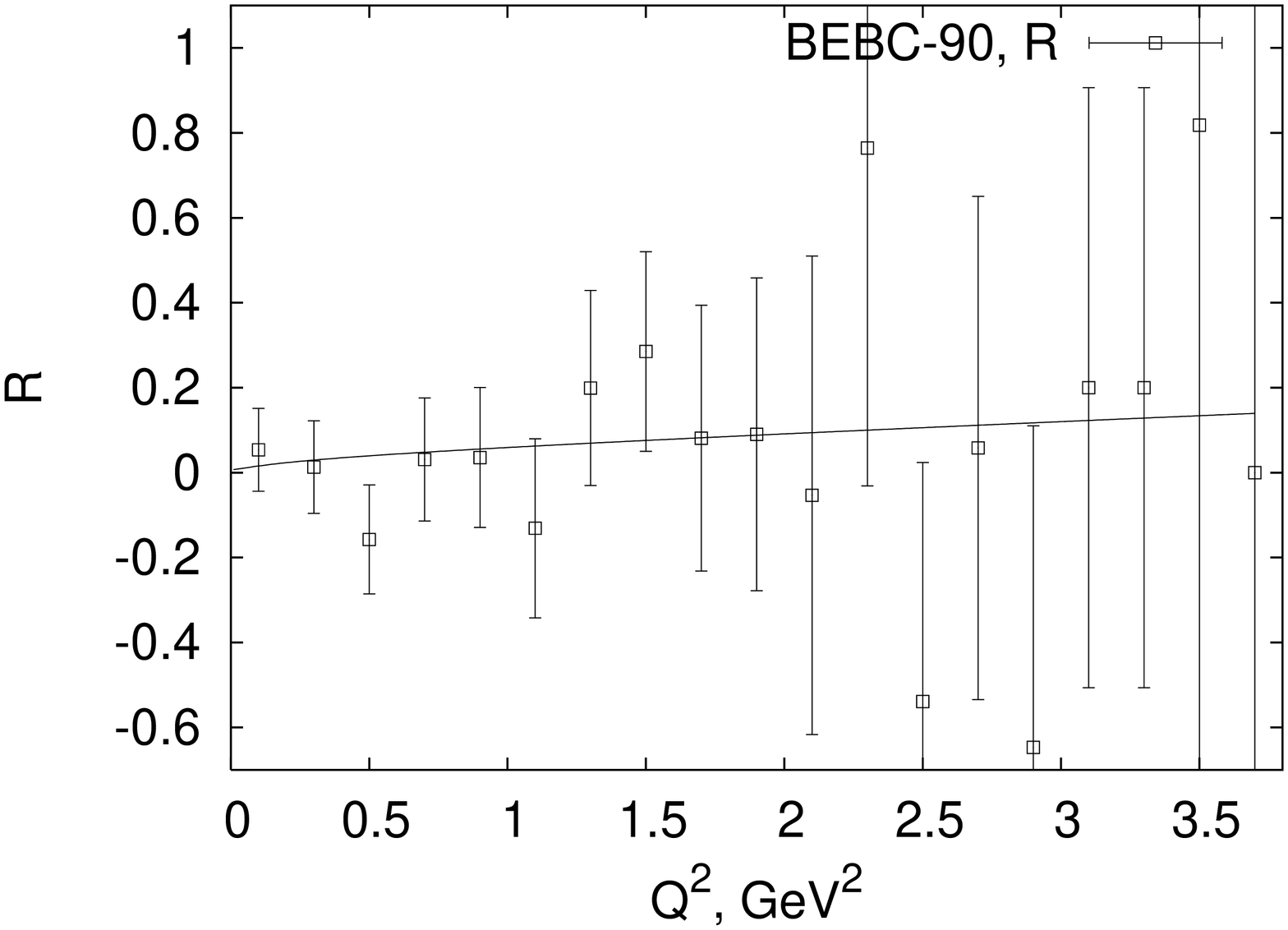}
\caption{Experiment BEBC-90: predictions for $R$.}
\label{bebc-90-}
\end{minipage}
\end{figure}

Comparison between experiments with neutrino and antineutrino beams will be interesting. The cross sections for $\nu p \to \mu^- p \pi^+$ and  $\bar\nu n \to \mu^+ n \pi^-$ are related to each other by changing the sign of the ${\cal W}_3$ term, i.e. the axial--vector interference term. It follows from Fig.~\ref{c3v5a}a, that the $Q^2-$distribution will be, at low energies, very different. Similarly the difference $\si_{\nu p}-\si_{\bar\nu n}$ is large at low energies and becomes smaller at high energies. This property follows from the fact, that the structure of the form factors limits the dominant contribution to the low region of $Q^2$.  Furthermore, the value of $q\cdot p=\nu m_N$ remains small in the resonance region. These two properties together imply that the contribution from ${\cal W}_2$ increases quadratically with $E_\nu$, while the vector--axial interference terms grow linearly. To illustrate this property, we define ratio

\beq
R=\frac{\di \frac{d \si_{\nu p}}{dQ^2} - \frac{d \si_{\bar\nu n}}{dQ^2}}{\di \frac{d \si_{\nu p}}{dQ^2} + \frac{d \si_{\bar\nu n}}{dQ^2}}
\label{R-Q2}
\eeq
and plot it in Fig.~\ref{pred-R} as a function of $Q^2$ for three energies. The theoretical curves terminate at values of $Q^2$ when the phase space is kinematically not allowed. 

The channel $\bar\nu n\to \mu^+ n \pi^-$ was observed in the BEBC-90 experiment for an average antineutrino energy $\langle E_{\bar\nu} \rangle=40\GeV$. Combining neutrino and antineutrino reactions we plot in Fig.~\ref{bebc-90-} the ration $R$ as a function of $Q^2$. For the high energies under consideration and the experimentally accessible region of $Q^2<3.5\GeV$ this ratio grows slowly from $0.0$ to $0.15$. 

\section{Summary and conclusions \label{summaryconclu}}
In this article we have calculated  general formulas for the production of $J=3/2$ resonances by neutrinos, including the muon mass. The formalism is applicable to the $P_{33}(1232)$, $D_{13}(1520)$ resonances provided the form factors are available. In this article we analyse the production of $P_{33}$ resonance and compare it with data that are available. The analysis of $D_{13}(1520)$ and the other $J=1/2$ resonances is left for a future publication.

Combining results from electroproduction and previous analysis of neutrino production \cite{Paschos:2003qr} we find again that the dominant contribution comes from the $C_3^V(Q^2)$ and $C_5^A(Q^2)$ form factors. Their dependence in $Q^2$ is faster than the dipole form factors.

A peculiar feature of the low energy data is a decrease of the differential cross section, $d\si / dQ^2$, as $Q^2\to 0$. We presented an analysis including the mass of the muon and Pauli blocking, both of which bring better agreement with the data. We have demonstrated that the mass of the muon restricts the phase space for the process in the low $Q^2$ region. The effects should be observable in the MiniBooNE and K2K experiments. In the small $Q^2$ region coherent scattering may also be present and should be established as a sharp peak in the $d\si/d Q^2 dt$ versus $t$ distribution, with the four-momentum-transfer squared given by 
\beq
t=(q-p_\pi)^2=-\left( \sum\limits_{\mu,\pi} p_i^{\perp} \right)^2 
              -\left( \sum\limits_{\mu,\pi}(E_i - p_i^{\parallel}) \right)^2
\label{t}
\eeq
This formula is based on zero energy transfer to the nucleus but includes the muon mass \cite{Rein:1982pf,Paschos:2003hs}. Then the incoherent sum of  the two effects must reproduce the data. 

It is still interesting to analyse other final state channels $p\pi^0$,  $n\pi^+$, \dots as well as reactions with antineutrino beams. In the low energy experiments, ${\cal W}_3(Q^2)$, which distinguishes neutrinos from antineutrinos, is large as it is shown in Fig.~\ref{W1W5}. For completeness we also presented results at higher energies where general trends are already apparent. For instance, the mass of the muon becomes less important at higher energies and the contribution from ${\cal W}_3(Q^2)$ is smaller.

A similar analysis is possible for the $D_{13}(1520)$ and $J=1/2$ resonances $P_{11}(1440)$, $S_{11}(1535)$, \dots, where information for the vector form factors is now available \cite{Burkert:2004sk,Aznauryan:2004jd} from Jefferson Lab. experiments. We plan to include this in the second article of this series.  

\begin{acknowledgements}
The financial support of BMBF, Bonn under contract  05HT 4 PEA/9 is  gratefully acknowledged. One of us (EAP) wishes to thank the theory group of Fermilab National Laboratory for ist hospitality where this work was completed. We are thankful to  Drs. C.~Albright, A.~Bodeck, S.~Zeller, H.~Gallagher and A.~Mann for helpful discussions and comments.
\end{acknowledgements}

\begin{widetext}

\appendix
\section{Structure of the hadronic tensor}

As we have mentioned, the hadronic tensor is parametrised  in the form (\ref{calW}).
The functions ${\cal W}_1, \dots, {\cal W}_6$ have been calculated from Eq.(\ref{wfordelta}) and led to 
\[
W_i(Q^2,\nu)=\frac{1}{m_N} V_i(Q^2,\nu) \delta(W^2-M_R^2)
\]
with the $V_i(Q^2,\nu)$ being the following
\beq
\begin{array}{l} \di
\frac{V_1}{3} = \frac{(C_3^V)^2}{m_N^2} \frac{2}{3 M_R^2}
                       \left[(\pq-Q^2)^2(\pq+m_N^2)+M_R^2((\pq)^2+Q^2 m_N^2+ Q^2 m_N M_R)\right]
\\[6mm] \hspace*{15mm} \di
+ \frac{(C_4^V)^2}{m_N^4} \frac{2}{3} (\pq-Q^2)^2(\pq+m_N^2-m_N M_R)
\\[6mm] \hspace*{15mm} \di

+ \frac{C_3^V C_4^V}{m_N^3} \frac{2}{3 M_R}
                          (\pq-Q^2) \left[ (\pq-Q^2)(\pq+m_N^2-2m_N M_R)+M_R^2 \pq \right]
\\[6mm] \hspace*{15mm} \di
+\frac{2}{3}\left[ 
\left(\frac{C_4^A}{m_N^2}\right)^2 (\pq-Q^2)^2  
+  (C_5^A)^2
+  2\frac{C_4^A C_5^A}{m_N^2} (\pq-Q^2) \right]
\left[  \pq+m_N^2+m_NM_R \right]
\end{array}
\label{calW1}
\eeq

\beq
\begin{array}{l} \di
\frac{V_2}{3} = 
(C_3^V)^2 \frac{2}{3M_R^2} Q^2 \left[ \pq +m_N^2 +M_R^2 \right]  
+
\frac{(C_4^V)^2}{m_N^2} \frac{2}{3} Q^2 \left[ \pq +m_N^2 - m_N M_R \right]  
\\[6mm] \hspace*{5mm} \di
+
\frac{C_3^V C_4^V}{m_N} \frac{2}{3M_R} Q^2 \left[ \pq + (M_R-m_N)^2 \right]  
+\frac23 \left[ 
({C_5^A})^2 \frac{m_N^2}{M_R^2} 
+ \frac{(C_4^A)^2}{m_N^2} Q^2 \right]
\left[  \pq+m_N^2+m_NM_R \right]
\end{array}
\label{calW2}
\eeq

\beq
\begin{array}{l} \di
\frac{V_3}{3} = \frac{4}{3M_R}\left[ -\frac{C_3^V C_4^A}{m_N}(\pq-Q^2)-C_3^V C_5^A m_N \right] 
\left[2M_R^2+2m_NM_R+Q^2-\pq \right]
\\[6mm] \hspace*{25mm} \di 
+ \frac43 (\pq - Q^2) \left[-\frac{C_4^V C_4^A}{m_N^2}(\pq-Q^2) -  C_4^V C_5^A  \right]
\end{array}
\label{calW3}
\eeq

These are the important form factors for most of the kinematic region. As mentioned already, there are two additional form factors, whose contribution to the cross section is proportional to the square of the muon mass.

\beq
\begin{array}{l} \di
\frac{V_4}{3}=
\frac{2}{3M_R^2} (C_3^V)^2 \left[ (2\pq-Q^2)(\pq+m_N^2) -M_R^2(m_N^2 +m_N M_R) \right]  
+
\frac{2}{3} \frac{(C_4^V)^2}{m_N^2}  (2\pq-Q^2) \left[ \pq +m_N^2 - m_N M_R \right]  
\\[6mm] \hspace*{15mm} \di
+
\frac{2}{3M_R} \frac{C_3^V C_4^V}{m_N}  \left[ (2\pq-Q^2)(\pq +m_N^2 - 2 m_N M_R)+ \pq M_R^2 \right]  
\\[6mm] \hspace*{15mm} \di
+\frac23 
\left[ 
(C_5^A)^2 \frac{m_N^2}{M_R^2}  
+ \frac{(C_4^A)^2}{m_N^2}(2\pq-Q^2) 
+\frac{(C_6^A)^2}{m_N^2 M_R^2}\left((Q^2-\pq)^2+Q^2M_R^2\right) 
\right.
\\[6mm] \hspace*{25mm} \di
\left.
+2 C_4^A C_5^A -2\frac{C_4^A C_6^A}{m_N^2}\pq -2\frac{C_5^A C_6^A}{M_R^2}(M_R^2+Q^2-\pq)
\right] 
\left[\pq +m_N^2 +m_N M_R \right]
\end{array}
\label{calW4}
\eeq

\beq
\begin{array}{l} \di
\frac{V_5}{3}=  
\frac23\frac{(C_3^V)^2}{M_R^2}\pq \left[ \pq +m_N^2+ M_R^2 \right]               
+\frac23 \frac{(C_4^V)^2}{m_N^2}\pq \left[ \pq +m_N^2 - m_N M_R \right]
\\[6mm] \hspace*{10mm} \di
+\frac2{3 M_R} \frac{C_3^V C_4^V}{m_N} \pq \left[ \pq + (M_R-m_N)^2 \right]
\\[6mm] \hspace*{10mm} \di
+\frac23 \left[ \frac{(C_4^A)^2}{m_N^2}\pq + (C_5^A)^2 \frac{m_N^2}{M_R^2}
                +C_4^A C_5^A  - \frac{C_4^A C_6^A}{m_N^2}Q^2 
               + \frac{C_5^A C_6^A}{M_R^2}(\pq - Q^2)
        \right] \left[ \pq +m_N^2+ m_N M_R \right]
\end{array}
\label{calW5}
\eeq

\beq
V_6=0
\label{calW6}
\eeq
\end{widetext}

Notice, as it is expected, for the contribution of the vector form factors the equalities
${\cal W}_5={\cal W}_2 \cdot (\pq) /Q^2$  
and  ${\cal W}_4={\cal W}_2 \cdot (\pq)^2 /Q^4 - {\cal W}_1 m_N^2 /Q^2$ are satisfied. 

In terms of the invariant variables, $Q^2$ and $W$, the scalar products of the 
4-vectors are:
\[
\begin{array}{c} \di
\pk=m_N E,
\quad 
\pq=m_N\nu,
\quad
\nu=\frac{W^2+Q^2-m_N^2}{2m_N} 
\\[7mm] \di
\pW=\frac12( W^2 + Q^2 + m_N^2 ),
\quad
\qW=m_N\nu-Q^2.
\end{array}
\]

Eq. (\ref{cs2}) must be compared with the known one from the Ref.\cite{Schreiner:1973mj}. This can be easily done for a specific case $Q^2\to 0$, $m_\mu\to 0$, when only $C_5^A$ contribute to the cross section. After the integration over $W$ with the help of the delta-function we obtain 
\[
\frac{d \si}{d Q^2}=\frac{G_F^2}{2\pi} \cos^2\theta_C 
(C_5^A)^2 
\left(1-\frac{M_R^2-m_N^2}{2 m_N E}\right) \frac{(m_N+M_R)^2}{2M_R^2},
\]
that identically coincides with the result, obtained in a similar way from Ref.\cite{Schreiner:1973mj}.

\bibliographystyle{apsrev}
\bibliography{references}

\end{document}